\documentstyle[12pt]{article}
\input epsf
\begin{document}
\setcounter{page}{0}
\thispagestyle{empty}
\vspace*{-54pt}
\begin{flushright}
{\footnotesize
FERMILAB--PUB--98/217--A\\
CERN-TH/98-227\\
hep-ph/9809453\\
September 1998\\}
\end{flushright}
\vspace*{0.5cm}
\begin{center}
{\Large \bf Production of massive particles during reheating}
\\
\vspace*{0.7cm}
{\bf
Daniel J. H. Chung,$^{a,b,}$\footnote{Electronic mail:
                      {\tt djchung@theory.uchicago.edu}}
Edward W. Kolb,$^{b,c,}$\footnote{Electronic mail:
                      {\tt  rocky@rigoletto.fnal.gov}}
and Antonio Riotto$^{d,}$\footnote{Electronic mail:
                      {\tt riotto@nxth04.cern.ch}}$^,$\footnote{On
                      leave  from Department of Theoretical Physics,
                      University of Oxford, U.K. }
}
\\
\vspace*{ 0.5cm}
{\it
$^a$Department of Physics and Enrico Fermi Institute \\
The University of Chicago, Chicago, Illinois 60637-1433\\
\vspace{12pt}
$^b$NASA/Fermilab Astrophysics Center \\ Fermilab
National Accelerator Laboratory, Batavia, Illinois~~60510-0500\\
\vspace{12pt}
$^c$Department of Astronomy and Astrophysics and
Enrico Fermi Institute\\
The University of Chicago, Chicago, Illinois~~60637-1433\\
\vspace{12pt}
$^d$Theory Division, CERN, CH-1211 Geneva 23, Switzerland
}
\end{center}
\vspace*{0.1cm}
\begin{quote}
\baselineskip=16pt
\hspace*{2em} What is commonly called the reheat temperature,
$T_{RH}$, is not the maximum temperature obtained after inflation.
The maximum temperature is, in fact, much larger than $T_{RH}$.  As an
application of this we consider the production of massive stable
dark-matter particles of mass $M_X$ during reheating, and show that
their abundance is suppressed as a power of $T_{RH}/M_X$ rather than
$\exp(-M_X/T_{RH})$.  We find that particles of mass as large as
$2\times 10^3$ times the reheat temperature may be produced in
interesting abundance.  In addition to dark matter, our analysis is
relevant for baryogenesis if the baryon asymmetry is produced by the
baryon (or lepton) number violating decays of superheavy bosons, and
also for relic ultra-high energy cosmic rays if decays of superheavy
particles are responsible for the highest energy cosmic rays.
\vspace*{4pt}

PACS number(s): 98.80.Cq

\end{quote}

\renewcommand{\thefootnote}{\arabic{footnote}}
\addtocounter{footnote}{-3}

\newpage

\setcounter{page}{1}

\def\simlt{\stackrel{<}{{}_\sim}}
\def\simgt{\stackrel{>}{{}_\sim}}

\baselineskip=24pt

\centerline{\bf I. INTRODUCTION}
\vspace{12pt}

At the end of inflation \cite{review} the energy density of the
universe is locked up in a combination of kinetic energy and potential
energy of the inflaton field, with the bulk of the inflaton energy
density in the zero-momentum mode of the field.  Thus, the universe at
the end of inflation is in a cold, low-entropy state with few degrees
of freedom, very much unlike the present hot, high-entropy universe.
After inflation the frozen inflaton-dominated universe must somehow be
defrosted and become a high-entropy radiation-dominated universe.

The process by which the inflaton energy density is converted to
radiation is known as ``reheating'' \cite{book}.  The possible role of
nonlinear dynamics leading to explosive particle production has
recently received a lot of attention.  This process, known as
``preheating'' \cite{kls} may convert a fair fraction of the inflaton
energy density into other degrees of freedom, with extremely
interesting cosmological effects such as symmetry restoration,
baryogenesis, or production of dark matter.  But the efficiency of
preheating is very sensitive to the model and the model parameters.
In some models the process is inefficient; in some models it is not
operative at all.  Even if preheating is relatively efficient, it is
unlikely to remove {\it all} of the energy density of in the inflaton
field.  It is likely that the slow decay of the inflaton field is
necessary to extract the remaining inflaton field energy.

The simplest way to envision this process is if the comoving energy
density in the zero mode of the inflaton decays into normal particles,
which then scatter and thermalize to form a thermal background.  It is
usually assumed that the decay width of this process is the same as
the decay width of a free inflaton field.

There are two reasons to suspect that the inflaton decay width might
be small.  The requisite flatness of the inflaton potential suggests a
weak coupling of the inflaton field to other fields since the
potential is renormalized by the inflaton coupling to other fields
\cite{review}.  However, this restriction may be evaded in
supersymmetric theories where the nonrenormalization theorem ensures a
cancelation between fields and their superpartners.  A second reason
to suspect weak coupling is that in local supersymmetric theories
gravitinos are produced during reheating.  Unless reheating is
delayed, gravitinos will be overproduced, leading to a large undesired
entropy production when they decay after big-bang nucleosynthesis
\cite{ellis}.

Of particular interest is a quantity known as the reheat temperature,
denoted as $T_{RH}$. The reheat temperature is calculated by assuming
an instantaneous conversion of the energy density in the inflaton
field into radiation when the decay width of the inflaton energy,
$\Gamma_\phi$, is equal to $H$, the expansion rate of the universe.

The reheat temperature is calculated quite easily \cite{book}.  After
inflation the inflaton field executes coherent oscillations about the
minimum of the potential.  Averaged over several oscillations, the
coherent oscillation energy density redshifts as matter: $\rho_\phi
\propto a^{-3}$, where $a$ is the Robertson--Walker scale factor.  If
we denote as $\rho_I$ and $a_I$ the total inflaton energy density and
the scale factor at the initiation of coherent oscillations, then the
Hubble expansion rate as a function of $a$ is ($M_{Pl}$ is the Planck
mass)
\begin{equation}
H(a) = \sqrt{\frac{8\pi}{3}\frac{\rho_I}{M^2_{Pl}}
	\left( \frac{a_I}{a} \right)^3}\ .
\end{equation}
Equating $H(a)$ and $\Gamma_\phi$ leads to an expression for $a_I/a$.
Now if we assume that all available coherent energy density is
instantaneously converted into radiation at this value of $a_I/a$, we
can define the reheat temperature by setting the coherent energy
density, $\rho_\phi=\rho_I(a_I/a)^3$, equal to the radiation energy
density, $\rho_R=(\pi^2/30)g_*T_{RH}^4$, where $g_*$ is the effective
number of relativistic degrees of freedom at temperature $T_{RH}$.
The result is
\begin{equation}
\label{eq:TRH}
T_{RH} = \left( \frac{90}{8\pi^3g_*} \right)^{1/4}
		\sqrt{ \Gamma_\phi M_{Pl} } \
       = 0.2 \left(\frac{200}{g_*}\right)^{1/4}
	      \sqrt{ \Gamma_\phi M_{Pl} } \ .
\label{eq:trh2}
\end{equation}
The limit from gravitino overproduction is $T_{RH} \simlt 10^{9}$ to
$10^{10}$ GeV.

The reheat temperature is best regarded as the temperature below which
the universe expands as a radiation-dominated universe, with the scale
factor decreasing as $g_*^{-1/3}T^{-1}$.  In this regard it has a
limited meaning \cite{book,turner}.  For instance, $T_{RH}$ {\em
should not} be used as the maximum temperature obtained by the
universe during reheating.  The maximum temperature is, in fact, much
larger than $T_{RH}$.  One implication of this is that it is incorrect
to assume that the maximum abundance of a massive particle species
produced after inflation is suppressed by a factor of
$\exp(-M/T_{RH})$.

In this paper we illustrate this effect by calculating the abundance
of a massive particle species produced in reheating.  We show that
particles of mass much greater than the eventual ``reheat''
temperature $T_{RH}$ may be created by the thermalized decay products
of the inflaton.  As an example, we demonstrate that a stable particle
species $X$ of mass $M_X$ would be produced in the reheating process
in sufficient abundance that its contribution to closure density today
is approximately $M_X^2 \langle \sigma |v|\rangle(g_*/200)^{-3/2}
(2000T_{RH}/M_X)^7$, where $g_*$ is the number of effective degrees of
freedom of the radiation energy density and $\langle \sigma|v|
\rangle$ is the thermal average of the $X$ annihilation cross section
times the M{\o}ller flux factor. Thus, particles of mass as large as
$2000$ times the reheat temperature may be produced in interesting
abundance.

Other applications of the effect include production of massive Higgs
bosons which could decay and produce the baryon asymmetry, or massive
particles that could decay and produce high-energy cosmic rays.

In the next section we develop a system of Boltzmann equations
describing the evolution of the energy densities of the inflaton
field, radiation, and a massive particle species.  In Section III we
find analytic approximations to the system and estimate the
contribution to the present critical density from a stable, massive
particle produced in reheating.  Section IV contains some numerical
results which illustrate several generic features.  We conclude in
Section V by discussing some applications of our results.  The
assumption of local thermodynamic equilibrium for the light degrees of
freedom is addressed in an appendix.

\vspace{36pt}
\centerline{\bf II. BOLTZMANN EQUATIONS DESCRIBING REHEATING}
\vspace{12pt}

Let us consider a model universe with three components: inflaton field
energy, $\rho_\phi$, radiation energy density, $\rho_R$, and the
energy density of a nonrelativistic particle species, $\rho_X$.  We
will assume that the decay rate of the inflaton field energy density
is $\Gamma_\phi$, with a branching fraction into $X\bar{X}$ of $B_X$,
and a branching fraction $1-B_X$ into light degrees of freedom,
generically referred to as radiation. We will denote the decay width
of the $X$ as $\Gamma_X$.  We will also assume that the light degrees
of freedom are in local thermodynamic equilibrium.  This is by no
means guaranteed, and we will return to the question in the appendix.

With the above assumptions, the Boltzmann equations describing the
redshift and interchange in the energy density among the different
components is
\begin{eqnarray}
\label{eq:BOLTZMANN}
& &\dot{\rho}_\phi + 3H\rho_\phi +\Gamma_\phi\rho_\phi = 0
	\nonumber \\
& & \dot{\rho}_R + 4H\rho_R - (1-B_X)\Gamma_\phi\rho_\phi
   - 	\frac{\langle\sigma|v|\rangle}{m_X}
	\left[ \rho_X^2 - \left( \rho_X^{EQ} \right)^2 \right]
   -  \Gamma_X  \left( \rho_X - \rho_X^{EQ}\right)= 0 \nonumber \\
& & \dot{\rho}_X + 3H\rho_X - B_X\Gamma_\phi \rho_\phi
    + \frac{\langle\sigma|v|\rangle}{m_X}
	\left[ \rho_X^2 - \left( \rho_X^{EQ} \right)^2 \right]
    + \Gamma_X  \left( \rho_X - \rho_X^{EQ} \right)
	= 0  \ ,
\end{eqnarray}
where dot denotes time derivative As already mentioned, $\langle
\sigma|v| \rangle$ is the thermal average of the $X$ annihilation
cross section times the M{\o}ller flux factor.  The equilibrium energy
density for the $X$ particles, $\rho_X^{EQ}$, is determined by the
radiation temperature, $T$.

It is useful to introduce two dimensionless constants, $\alpha_\phi$
and $\alpha_X$, defined in terms of $\Gamma_\phi$ and $\langle \sigma
|v| \rangle$ as
\begin{equation}
\label{alphagamma}
\Gamma_\phi = \alpha_\phi M_\phi \qquad
\langle \sigma |v| \rangle = \alpha_X M_X^{-2} \ .
\end{equation}
For a reheat temperature much smaller than $M_\phi$, $\Gamma_\phi$
must be small.  From Eq.\ (\ref{eq:TRH}), the reheat temperature in
terms of $\alpha_X$ and $M_X$ is $T_{RH}\simeq \alpha_\phi^{1/2}
\sqrt{M_\phi M_{Pl}}$.  For $M_\phi=10^{13}$GeV, $\alpha_\phi$ must be
smaller than of order $10^{-13}$.  On the other hand, $\alpha_X$ may
be as large as of order unity, or it may be small also.

In what follows we will make the simplifying assumption that $B_X=0$.
Since we are interested in a stable particle relic, we will assume
that $\Gamma_X=0$.  It is also convenient to work with dimensionless
quantities that can absorb the effect of expansion of the universe.
This may be accomplished with the definitions
\begin{equation}
\label{def}
\Phi \equiv \rho_\phi M_\phi^{-1} a^3 \ ; \quad
R    \equiv \rho_R a^4 \ ; \quad
X    \equiv \rho_X M_X^{-1} a^3 \ .
\end{equation}
It is also convenient to use the scale factor, rather than time, for
the independent variable, so we define a variable $x = a M_\phi$.
With this choice the system of equations can be written as (prime
denotes $d/dx$)
\begin{eqnarray}
\label{eq:SYS}
\Phi' & = & - c_1 \ \frac{x}{\sqrt{\Phi x + R}}   \ \Phi \nonumber \\
R'    & = &   c_1 \ \frac{x^2}{\sqrt{\Phi x + R}} \ \Phi \
            + c_2 \ \frac{x^{-1}}{ \sqrt{\Phi x +R}} \
	           	         \left( X^2 - X_{EQ}^2 \right) \nonumber \\
X'    & = & - c_3 \ \frac{x^{-2}}{\sqrt{\Phi x +R}} \
		\left( X^2 - X_{EQ}^2 \right) \ .
\end{eqnarray}
The constants $c_1$, $c_2$, and $c_3$ are given by
\begin{equation}
c_1 = \sqrt{\frac{3}{8\pi}} \frac{M_{Pl}}{M_\phi}\alpha_\phi \ \qquad
c_2 = c_1\frac{M_\phi}{M_X}\frac{\alpha_X}{\alpha_\phi} \ \qquad
c_3 = c_2 \frac{M_\phi}{M_X} \ .
\end{equation}
$X_{EQ}$ is the equilibrium value of $X$, given in terms of the
temperature $T$ as (assuming a single degree of freedom for the $X$
species)
\begin{equation}
X_{EQ} = \frac{M_X^3}{M_\phi^3}\left( \frac{1}{2\pi} \right)^{3/2}
	x^3 \left(\frac{T}{M_X}\right)^{3/2}\exp(-M_X/T) \ .
\end{equation}
The temperature depends upon $R$ and $g_*$, the effective number of
degrees of freedom in the radiation:
\begin{equation}
\frac{T}{M_X} = \left( \frac{30}{g_*\pi^2}\right)^{1/4}
\frac{M_\phi}{M_X} \frac{R^{1/4}}{x} \ .
\end{equation}

It is straightforward to solve the system of equations in Eq.\
(\ref{eq:SYS}) with initial conditions at $x=x_I$ of $R(x_I)=X(x_I)=0$
and $\Phi(x_I)=\Phi_I$.  It is convenient to express
$\rho_\phi(x=x_I)$ in terms of the expansion rate at $x_I$, which
leads to
\begin{equation}
\Phi_I = \frac{3}{8\pi} \frac{M^2_{Pl}}{M_\phi^2}
		\frac{H_I^2}{M_\phi^2}\ x_I^3 \ .
\end{equation}
The numerical value of $x_I$ is irrelevant.

Before numerically solving the system of equations, it is useful to
consider the early-time solution for $R$.  Here, by early time, we
mean $H \gg \Gamma_\phi$, i.e., before a significant fraction of the
comoving coherent energy density is converted to radiation.  At early
times $\Phi \simeq \Phi_I$, and $R\simeq X \simeq 0$, so the equation
for $R'$ becomes $R' = c_1 x^{3/2} \Phi_I^{1/2}$.  Thus, the early
time solution for $R$ is simple to obtain:
\begin{equation}
\label{eq:SMALLTIME}
R \simeq \frac{2}{5} c_1
     \left( x^{5/2} -  x_I^{5/2} \right) \Phi_I^{1/2}
			 \qquad (H \gg \Gamma_\phi) \ .
\end{equation}
Now we may express $T$ in terms of $R$ to yield the early-time
solution for $T$:
\begin{equation}
\label{threeeights}
\frac{T}{M_\phi} \simeq \left(\frac{12}{\pi^2g_*}\right)^{1/4}
c_1^{1/4}\left(\frac{\Phi_I}{x_I^3}\right)^{1/8}
	\left[ \left(\frac{x}{x_I}\right)^{-3/2} -
                \left(\frac{x}{x_I}\right)^{-4} \right]^{1/4}
		\qquad (H \gg \Gamma_\phi) \ .
\label{eq:approxtovmphi}
\end{equation}
Thus, $T$ has a maximum value of
\begin{eqnarray}
\frac{T_{MAX}}{M_\phi}& = & 0.77
   \left(\frac{12}{\pi^2g_*}\right)^{1/4} c_1^{1/4}
   \left(\frac{\Phi_I}{x_I^3}\right)^{1/8} \nonumber \\ & = & 0.77
   \alpha_\phi^{1/4}\left(\frac{9}{2\pi^3g_*}\right)^{1/4} \left(
   \frac{M_{Pl}^2H_I}{M_\phi^3}\right)^{1/4} \ ,
\end{eqnarray}
which is obtained at $x/x_I = (8/3)^{2/5} = 1.48$.  It is also
possible to express $\alpha_\phi$ in terms of $T_{RH}$ and obtain
\begin{equation}
\label{max}
\frac{T_{MAX}}{T_{RH}} = 0.77 \left(\frac{9}{5\pi^3g_*}\right)^{1/8}
		\left(\frac{H_I M_{Pl}}{T_{RH}^2}\right)^{1/4} \ .
\end{equation}

For an illustration, in the simplest model of chaotic inflation $H_I
\sim M_\phi$ with $M_\phi \simeq 10^{13}$GeV, which leads to
$T_{MAX}/T_{RH} \sim 10^3 (200/g_*)^{1/8}$ for $T_{RH} =
10^9$GeV.

We can see from Eq.\ (\ref{eq:SMALLTIME}) that for $x/x_I>1$, in the
early-time regime $T$ scales as $a^{-3/8}$, which implies that entropy
is created in the early-time regime \cite{turner}.  So if one is
producing a massive particle during reheating it is necessary to take
into account the fact that the maximum temperature is greater than
$T_{RH}$, and that during the early-time evolution, $T\propto
a^{-3/8}$.

\vspace{36pt}
\centerline{\bf III. PRODUCTION OF A MASSIVE, STABLE PARTICLE SPECIES}
\vspace{12pt}

\centerline{{\bf A. Freeze out of the comoving $X$ energy density}}
\vspace{12pt}

In this section we develop the equation for the $X_F$, the final value
of $X$, which can be found from the early-time behavior.

At early times $\Phi \simeq \Phi_I$ and $R \simeq 0$.  We will here
also assume that $X \ll X_{EQ}$.  Numerical results confirm the
validity of this approximation and show that the massive particles are
never in chemical equilibrium (although presumably they are in kinetic
equilibrium).  The early-time equation for the development of the $X$
energy density is
\begin{equation}
\label{noback}
X^\prime = c_3 \Phi_I^{-1/2} x^{-5/2} X^2_{EQ} \ .
\label{eq:xdevelop}
\end{equation}
$X_{EQ}$ is given in terms of $M_X$ and the temperature, which may be
found from the early-time solution for $R$.

We can integrate Eq.\ (\ref{eq:xdevelop}) by approximating it as a
Gaussian integral.  First we rearrange Eq.\ (\ref{eq:xdevelop}) by
making appropriate redefinitions.  Define the quantities $y$ and $\nu$
by $y \equiv X/x_I^3$ and $\nu \equiv (x/x_I)^{3/16}$.  Now, using
Eq.\ (\ref{eq:approxtovmphi}), we can rewrite Eq.\ (\ref{eq:xdevelop})
as
\begin{equation}
y(\nu)
= Q \int_1^{\nu} \! d\nu' \, \exp[- H(\nu') ] \,
\end{equation}
 where we have defined
\begin{eqnarray}
Q & \equiv & \alpha_\phi^{3/4} \alpha_X\left (\frac{3^{1/2}2^{1/4}}
{\pi^{21/4} g_*^{3/4}}\right) \frac{M_{Pl}^{3/2} M_X}{H_I^{1/4}
M_\phi^{9/4}} \nonumber \\ H(\nu) & \equiv & \lambda \left(\nu^{-8}
-\nu^{-64/3}\right)^{-1/4} - \frac{3}{4} \ln\left(\nu^{-8} -
\nu^{-64/3}\right) - 23 \ln{\nu} \nonumber \\ 
\lambda & \equiv & \frac{2^{5/4}\pi^{3/4}g_*^{1/4}}{3^{1/2}}
\frac{\alpha_\phi^{-1/4}M_X}{M_\phi^{1/4} H_I^{1/4}M_{Pl}^{1/2}} \ .
\end{eqnarray}

To proceed with the Gaussian integral approximation, we assume
$\nu_0^{-8} \gg \nu_0^{-64/3}$ where $\nu_0$ is the solution to
$H'(\nu_0)=0$.  Then, we can easily solve $H'(\nu_0)=0$, finding
$\nu_0 = \sqrt{17/2\lambda}$, which is the point about which we Taylor
expand $H(\nu)$ to quadratic order.  Since the integrand falls to 0
rapidly away from $\nu=\nu_0$, and since we desire the freeze out
value for $y$, the limits of the integrand can be taken to $\pm
\infty$.  We thereby find
\begin{equation}
\frac{X_F}{x_I^3} \approx y_\infty
\approx \frac{3^5\alpha_\phi^3\alpha_X} {8 \pi^{23/2} g_*^3}
\left(\frac{H_I^2 M_{Pl}^6}{M_X^8 }\right)
\left(\frac{\sqrt{17}}{2}\right)^{17} \exp(-17/2).
\end{equation}
 Using Eq.\ (\ref{eq:trh2}), we rewrite this in a more transparent form
\begin{equation}
\label{bb}
\frac{X_F}{x_I^3} \approx \frac{4.21\times10^{-6}}{(g_*/200)^{3/2}} \
\frac{\alpha_X H_I^2 M_{Pl}^3 T_{RH}^6}{M_\phi^3 M_X^8} \ .
 \end{equation}

Note that this approximation should be valid as long as $\nu_0^{-8}
\gg \nu_0^{-64/3}$ is satisfied.  If the condition is not satisfied,
the suppression will be exponential in $M_X/T$.

\vspace{24pt}
\centerline{{\bf B. $\Omega_Xh^2$ in terms of $X_F$}}
\vspace{12pt}

After freeze out of the comoving energy density of the stable
particle, $X$ remains constant, so $\rho_X(x>x_F) = X_F x_I^{-3} M_X
M^3_\phi (x_I/x)^3$.  For delayed reheating ($\Gamma_\phi \ll H_I$)
freeze out will be well before reheating.  After reheating,
$\rho_X(x>x_{RH}) = \rho_X(x_{RH})(x_{RH}/x)^3$.  The comoving entropy
density is constant after reheating, so the radiation energy density
scales as $\rho_R(x>x_{RH}) = \rho_R(x_{RH})
[g_*(T_{RH})/g_*(T)]^{1/3} (x_{RH}/x)^4$.  Using these facts, we can
express the present contribution of the massive particle species to
the critical density in terms of the ratio of the energy densities at
freeze out:
\begin{equation}
\frac{\Omega_X h^2}{\Omega_R h^2}
  =	\frac{\rho_X(T_{RH})}{\rho_R(T_{RH})}
  	\left(\frac{g_*(\mbox{today})}{g_*(T_{RH})}\right)^{1/3}
	\frac{x_0}{x_{RH}}
  =  	\frac{\rho_X(T_{RH})}{\rho_R(T_{RH})}\frac{T_{RH}}{T_0}
		\quad \qquad (\mbox {for}\ x>x_{RH}) \ ,
\end{equation}
where $x_0$ is the present value of $x$ and $T_0 = 2.37 \times
10^{-13}$GeV is the present temperature.\footnote{In this subsection
we make the heretofore criticized approximation that the inflaton
energy density scales like pressureless matter until it dumps all of
its energy into radiation at the instant of ``reheating.''  In this
instance, however, it is an appropriate approximation, as borne out by
analytic approximations and the numerical calculations presented in
the next section.}  Today, $\Omega_R h^2 = 4.3\times 10^{-5}$, and the
contribution to $\Omega h^2$ from the massive particle is
\begin{equation}
\Omega_Xh^2 = 1.5 \times 10^{18}
	\left(\frac{T_{RH}}{10^9{\rm GeV}}\right)
	\frac{X_F}{x_I^3}\frac{M_X M^3_\phi}{H^2_I M^2_{Pl}} \ .
\end{equation}
Using the expression for $X_F/x_I^3$ from the previous section, we
arrive at the final result
\begin{equation}
\label{om}
\Omega_X h^2 = M_X^2 \langle \sigma |v|\rangle \,
	\left(\frac{g_*}{200}\right)^{-3/2} \,
	\left (\frac{2000T_{RH}}{M_X}\right)^7 \ .
\end{equation}

\vspace{36pt}
\centerline{\bf IV. NUMERICAL RESULTS}
\vspace{12pt}

\begin{figure}[t]
\centering
\leavevmode\epsfxsize=400pt  \epsfbox{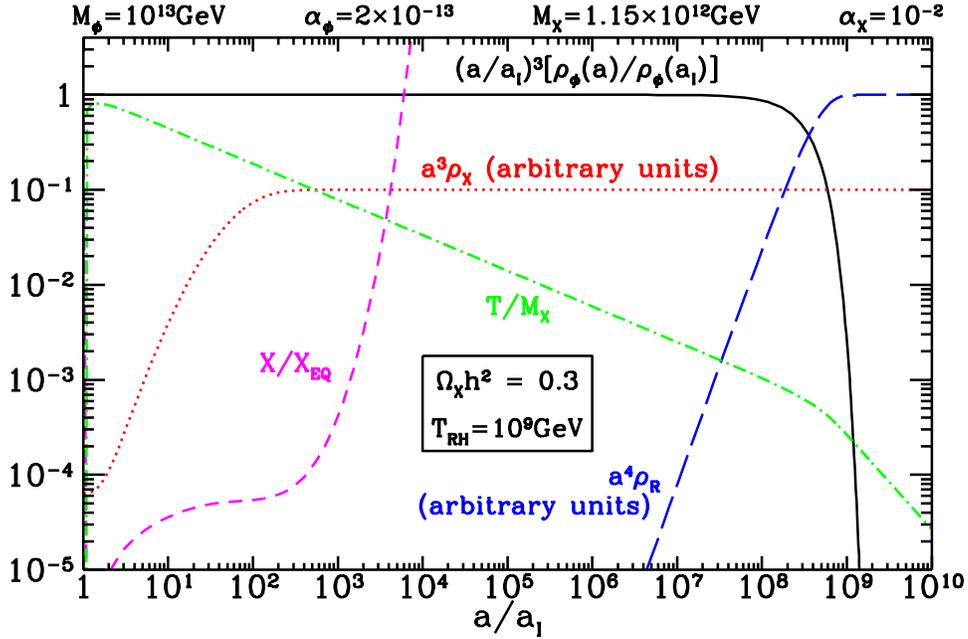}
\caption[fig1]{\label{model1}
The evolution of energy densities and $T/M_X$ as a function
of the scale factor.   Also shown is $X/X_{EQ}$.}
\end{figure}

An example of a numerical evaluation of the complete system in Eq.\
(\ref{eq:SYS}) is shown in Fig.\ \ref{model1}.  The model parameters
chosen were $M_\phi= 10^{13}$GeV, $\alpha_\phi =2\times10^{-13} $,
$M_X= 1.15\times10^{12}$GeV, $\alpha_X =10^{-2}$, and $g_*=200$.  The
expansion rate at the beginning of the coherent oscillation period was
chosen to be $H_I=M_\phi$.  These parameters result in
$T_{RH}=10^9$GeV and $\Omega_Xh^2=0.3$.

Figure \ref{model1} serves to illustrate several aspects of the
problem.  Just as expected, the comoving energy density of $\phi$
(i.e., $a^3\rho_\phi$) remains roughly constant until
$\Gamma_\phi\simeq H$, which for the chosen model parameters occurs
around $a/a_I\simeq 5\times10^8$.  But of course, that does not mean
that the temperature is zero.  Notice that the temperature peaks well
before ``reheating.''  The maximum temperature, $T_{MAX}= 10^{12}$GeV,
is reached at $a/a_I$ slightly larger than unity (in fact at
$a/a_I=1.48$ as expected), while the reheat temperature, $T_{RH}=
10^9$GeV, occurs much later, around $a/a_I\sim 10^8$.  Note that
$T_{MAX}\simeq 10^3 T_{RH}$ in agreement with Eq.\ (\ref{max}).

{}From the numerical results we can justify one of the assumptions in
deriving the analytical approximations. From the figure it is clear
that $X \ll X_{EQ}$ at the epoch of freeze out of the comoving $X$
number density, which occurs around $a/a_I\simeq 10^2$.  The rapid rise
of the ratio after freeze out is simply a reflection of the fact that
$X$ is constant while $X_{EQ}$ decreases exponentially.  The relevance
of the ratio is the justification of the neglect of $X_{EQ}$ term in
Eq.\ (\ref{noback}).

A close examination of the behavior of $T$ shows that after the sharp
initial rise of the temperature, the temperature decreases as
$a^{-3/8}$ [as follows from Eq.\ (\ref{threeeights})] until $H\simeq
\Gamma_\phi$, and thereafter $T\propto a^{-1}$ as expected for the
radiation-dominated era.

For the choices of $M_\phi$, $\alpha_\phi$, $g_*$, and $\alpha_X$ used for
the model illustrated in Fig.\ \ref{model1}, $\Omega_Xh^2 = 0.3$ for  
$M_X=1.15\times10^{12}$GeV, in excellent agreement with the mass predicted 
by using Eq.\ (\ref{om}).  

\vspace{36pt}
\centerline{\bf V. CONCLUSIONS}
\vspace{12pt}

Let us now analyze the implications of our findings for the GUT
baryogenesis scenario, where the baryon asymmetry is produced by the
baryon (or lepton) number violating decays of superheavy bosons
\cite{bau}.  At the end of inflation the Universe does not contain any
matter and, even more important, it is perfectly baryon
symmetric---there is no dominance of matter over antimatter. This
means that GUT baryogenesis may be operative only if the supermassive
GUT bosons are regenerated during the stage of thermalization of the
decay products of the inflaton field\footnote{In the case in which
reheating is anticipated by a period of preheating, superheavy
bosons may be produced by the phenomenon of 
parametric resonance \cite{gutpre}.}. 
A naive estimate would lead to
conclude that that the maximum number density of a massive particle
species $X$ produced after inflation is suppressed by a factor of
$(M_X/T_{RH})^{3/2}\exp(-M_X/T_{RH})$ with respect to the photon
number density. For such a reason, it is commonly believed that GUT
baryogenesis is incompatible with models of inflation where the
reheating temperature is much smaller than the GUT scale and, in
general, than the mass of the $X$ particles, $T_{RH}\ll M_X$. In fact,
we have seen that the reheat temperature has a limited meaning and
should not be used as the maximum temperature obtained by the universe
during reheating.  The maximum temperature, Eq.\ (\ref{max}), is much
larger than $T_{RH}$, and particles of mass much greater than the
eventual reheating temperature $T_{RH}$ may be created by the
thermalized decay products of the inflaton {\it without} any
exponential suppression factor.  Indeed, the number density $n_X$ of
particles $X$ after freeze out and reheating may be easily inferred
from Eqs.\ (\ref{def}) and (\ref{bb}), and reads
\begin{equation}
\label{ph}
\frac{n_X}{n_\gamma}\simeq 3\times
10^{-4}\left(\frac{100}{g_*}\right)^{3/2}
\left(\frac{T_{RH}}{M_X}\right)^7\left(\frac{M_{Pl}}{M_X}\right).
\end{equation}

In theories that invoke supersymmetry to preserve the flatness of the
inflaton potential, the slow decay rate of the gravitinos, the
superpartners of the gravitons, is a source of the cosmological
problems because the decay products of the gravitino will destroy the
$^4$He and D nuclei by photodissociation, and in the process destroy
the successful nucleosynthesis predictions. The most stringent bound
comes from the resulting overproduction of D $+$ $^3$He, which would
require that the gravitino abundance is smaller than about $10^{-10}$
relative to the entropy density at the time of reheating after
inflation.  This translates into an upper bound on the reheating
temperature after inflation, $T_{RH}/M_{Pl}\simlt 10^{-9}$ \cite{ellis}.

It is easy to check that for such small values of $T_{RH}$ the ratio
in Eq.\ (\ref{ph}) is always much larger than the equilibrium value,
$n_X^{EQ}/n_\gamma=(M_X/2T_{RH})^{3/2}(\pi^{1/2}/\xi(3))\:{\rm
exp}(-M_X/T_{RH})$. This result is crucial for the out-of-equilibrium
decay scenarios of baryogenesis. For instance, in theories where $B-L$
is a spontaneously broken local symmetry, as suggested by $SO(10)$
unification, the cosmological baryon asymmetry can be generated by the
out-of-equilibrium decay of the lightest heavy Majorana right-handed
neutrino $N_1^c$ \cite{fu}, whose typical mass is about $10^{10}$ GeV.
For reheat temperatures of the order of $10^9$ GeV, the number density
of the right-handed neutrino is about $3\times 10^{-2}\:n_\gamma$ and
one can estimate the final baryon number to be of the order of $B\sim
(n_{N_1^c}/n_\gamma)(\epsilon/g_*)\simeq 10^{-4}\epsilon$, where
$\epsilon$ is the coefficient containing one-loop suppression factor
and $CP$ violating phases. The observed value of the baryon asymmetry,
$B\sim 10^{-10}$, is then obtained without any fine tuning of
parameters.

Our findings have also important implications for the conjecture that
ultra-high cosmic rays, above the Greisen-Zatsepin-Kuzmin cut-off of
the cosmic ray spectrum, may be produced in decays of superheavy
long-living particles \cite{kr1,kr2}. In order to produce cosmic rays
of energies larger than about $10^{13}$ GeV, the mass of the
$X$-particles must be very large, $M_X\simgt 10^{13}$ GeV and their
lifetime $\tau_X$ cannot be much smaller than the age of the Universe,
$\tau_X\simgt 10^{10}$ yr.  With the smallest value of the lifetime,
the observed flux of ultra-high energy cosmic rays will be reproduced
with a rather low density of $X$-particles, $\Omega_X\sim
10^{-12}$. It has been suggested that $X$-particles can be produced in
the right amount by usual collisions and decay processes taking place
during the reheating stage after inflation, if the reheat temperature
never exceeded $M_X$ \cite{kr2}.  Again, assuming naively that that
the maximum number density of a massive particle species $X$ produced
after inflation is suppressed by a factor of
$(M_X/T_{RH})^{3/2}\exp(-M_X/T_{RH})$ with respect to the photon
number density, one concludes that the reheat temperature $T_{RH}$
should be in the range $10^{11}$ to $10^{15}$GeV \cite{kr1}. This is a
rather high value and leads to the gravitino problem in generic
supersymmetric models.  This is one reason alternative production
mechanisms of these superheavy $X$-particles have been proposed
\cite{ckr1,kt,ckr2}. However, our analysis show that the situation is
much more promising. Making use of Eq.\ (\ref{om}), the right amount
of $X$-particles to explain the observed ultra-high energy cosmic rays
is produced for
\begin{equation}
\left(\frac{T_{RH}}{10^{10}\:{\rm GeV}}\right)\simeq
\left(\frac{g_*}{200}\right)^{3/14}\:\left(\frac{M_X}{10^{15}\:{\rm
GeV}}\right),
\end{equation}
where we have assumed $\langle \sigma |v|\rangle\sim
M_X^{-2}$. Therefore, we conclude that particles as massive as
$10^{15}$ GeV may be generated during the reheating stage in
abundances large enough to explain the ultra-high energy cosmic rays
even if the reheat temperature satisfies the gravitino bound.

\newpage
\vspace{36pt}
\centerline{\bf ACKNOWLEDGEMENTS}
\vspace{12pt}

DJHC and EWK were supported by the DOE and NASA under Grant NAG5-7092.

\vspace{36pt}
\centerline{\bf APPENDIX A: THERMALIZATION OF LIGHT DEGREES OF FREEDOM}
\vspace{12pt}
\renewcommand\theequation{A\arabic{equation}}
\setcounter{equation}{0}

The form of the Boltzmann equations we use, e.g., Eq.\
(\ref{eq:BOLTZMANN}), assumes that the light particle decay products
of the inflaton field are in local thermodynamic equilibrium (LTE).
In this appendix we discuss this assumption, and the implications if
it is not valid.

Before discussing the validity of the assumption, it is useful to
recall why the assumption was made.  In the derivation of Eq.\
(\ref{eq:BOLTZMANN}), one starts with an equation for the rate of
change of the $X$ number density due to the process
$\gamma\gamma\rightarrow XX$ with four-momentum conservation $p_\gamma
+ p^\prime_\gamma = p_X + p^\prime_X$:
\begin{eqnarray}
\dot{n}_X & = &  \int \frac{d^3\!p_\gamma}{2E_\gamma} 
                 \int \frac{d^3\!p^\prime_\gamma}{2E^\prime_\gamma}
	         \int \frac{d^3\!p_X}{2E_X} 
                 \int \frac{d^3\!p^\prime_{X}}{2E^\prime_{X}}
	         \left(2\pi\right)^{-8} 
      \delta^4\left(p_\gamma + p^\prime_\gamma - p_X - p^\prime_X  \right)
			    \nonumber \\
		  &    & \times
                  f_\gamma(p_\gamma)f_\gamma(p^\prime_\gamma)
	\left|{\cal M}\right|^2_{\gamma\gamma\rightarrow XX} +\cdots  \ .
\label{phasespace}
\end{eqnarray}
Here $f_i(p)$ is the phase-space density of particle species $i$ with
momentum $p_i$, and $\left|{\cal M}\right|^2_{\gamma\gamma\rightarrow
XX} $ is the square of the matrix element for the process
$\gamma\gamma\rightarrow XX$.  With the assumption that the light
particles are in LTE, the product of the light particle phase-space
densities is $f_\gamma(p_\gamma)f_\gamma(p^\prime_\gamma) =
\exp(-E_\gamma/T) \exp(-E^\prime_\gamma/T)$.  This last product is, of
course, simply $f_X^{EQ}(p_X) f_X^{EQ}(p^\prime_X)$, which, after some
rearrangement of Eq.\ (\ref{phasespace}), leads to a term for the
creation of $X$'s proportional to $\left(n^{EQ}_X\right)^2$:
\begin{equation}
\label{eq}
\dot{n}_X = \langle \sigma |v| \rangle \left(n^{EQ}_X\right)^2 + \cdots \ .
\end{equation}
(for complete details, see \cite{book}).   

The factor $\left( n^{EQ}_X \right)^2$ in Eq.\ (\ref{eq}) is present because 
not every light-particle collision has sufficient center-of-mass energy to
create an $X$ pair.  If LTE is established with temperature $T<M_X$,
the factor $\exp(-2M_X/T)$ in $\left(n^{EQ}_X\right)^2$ represents the
fraction of the collisions with center-of-mass energy above threshold, i.e., 
with $\sqrt{s} > 2M_X$.

A simple indication of whether thermalization occurs on a timescale
shorter than the timescale for $X$ production is the ratio of the
cross section for the thermalization reactions to the cross section
for $X$ production.  If the ratio is larger than unity, then
thermalization of the light degrees of freedom is a good assumption.

The process of $X$ production involves a ``hard'' process, and the
cross section will be $\alpha_X/M_X^2$, where $\alpha_X$ was defined in
Eq.\ (\ref{alphagamma}).  In order to produce an equilibrium
distribution from the original decay distribution it is necessary to
change the number of particles.  Therefore, the relevant cross section
is the one for processes like $\gamma\gamma \rightarrow
\gamma\gamma\gamma$.\footnote{Recall that $\gamma$ represents a light
particle, not just a photon, so $\gamma$ may carry electric charge,
color charge, etc.}  Although the thermalization reaction is higher
order in perturbation theory, it is a ``softer'' process, and
radiation of a soft photon has a large cross section.

Without knowing the details of the interactions of the decay products,
it is impossible to say with certainty how complete thermalization
will be.  But if the inflaton decay products have usual gauge
interactions, the thermalization cross section will be larger than the
$X$ production cross section, and thermalization of the inflaton decay
products is likely.

Now let's explore the consequences if LTE of the light degrees of
freedom is not obtained.  If the light particles are not in LTE, then
the factor $n^{EQ}_X$ in Eq.\ (\ref{eq}) could simply be replaced by
the more general factor $n_\gamma(E>M_X)$.\footnote{Of course in this
case $\langle \sigma |v|\rangle$ would not be a thermal average, but
an average over the actual phase-space density. } Now let's make the
extreme assumption that the light degrees of freedom {\it never}
interact before $X$ production, and that they have the original
(redshifted) momentum with which they were created in inflaton decay.
Assume this original momentum is $M_\phi/\eta$.  If $M_\phi/\eta$ is
greater than $M_X$, then the $X$ production rate actually will be
larger than the equilibrium rate since $n_\gamma(E>M_X) > n^{EQ}_X$,
while if $M_\phi/\eta$ is less than $M_X$, then the $X$ production
rate will be zero!  The most reasonable assumption is that even if
there is a large multiplicity in $\phi$ decay, a fair fraction of
$M_\phi$ is carried by a few leading particles.  So the effective
value of $\eta$ is probably not too large.

The above analysis leads us to the conclusion that thermalization of the
light degrees of freedom is likely unless the inflaton decay products
themselves are very weakly coupled to everything.  Even if
thermalization does not occur, production of massive particles during
reheating is not much different than our simple model suggests.


\end{document}